\documentclass[a4paper,12pt]{article}
\usepackage[utf8]{inputenc}

\usepackage{amsmath,amssymb}
\usepackage{graphicx}
\usepackage{textcomp}

\title{How to reduce the crack density in drying colloidal material?}
\author{Fran\c{c}ois Boulogne\textsuperscript{1,2}, Fr\'ed\'erique Giorgiutti-Dauphin\'e\textsuperscript{1}, Ludovic Pauchard\textsuperscript{1}}

\begin{document}

\maketitle
\emph{\textsuperscript{1} UPMC Univ Paris 06, Univ Paris-Sud, CNRS, F-91405. Lab FAST, Bat 502, Campus Univ, Orsay, F-91405, France.\\
\textsuperscript{2} boulogne@fast.u-psud.fr 
}

\begin{abstract}
 The drying of a colloidal dispersion can result in a gel phase defined as a porous matrix saturated in solvent.
 During the drying process, high mechanical stresses are generated.
When these stresses exceed the strength of the material, they can be released in the formation of cracks.
This process strongly depends on both the mechanical properties of the material and the way the gel consolidates.
In this report, we give experimental evidences that the number of cracks formed in the consolidating film depend on the drying rate, the nature of the solvent and the mechanical properties of the colloidal particles.
\end{abstract}

\section{Introduction}
\label{intro}
Coatings are usually made through the deposition of a volatile liquid that contains dispersed colloidal particles.
The dry coating is obtained through evaporation of the volatile liquid.
However, crack patterns often affect the final quality of the coating and often need to be avoided.
Numerous studies on cracks induced by drying of colloidal suspensions have been investigated for the last two decades.
Nevertheless the relation between the stress release and the resulting crack pattern remains unclear.
Particularly, during the first stage of the drying process, particles concentrate until the formation of a porous matrix saturated with solvent .
Capillary shrinkage at the evaporation surface, limited by adhesion on a substrate, results in sufficiently high tensile stress able to compete with the material cohesion (ie critical stress for fracture).
When the stress reaches a threshold value, the material responds by a re-organisation of its structure at different scales in order to release the increasing stress.
At the mesoscopic scale, the stress release can lead to the formation of various spatial patterns, like cracks.
The resulting crack patterns depend notably on the following parameters:

\begin{itemize}
\item the thickness of the layer in which the elastic energy of the system is stored\cite{Lazarus11},
\item the permeability of the  porous media, e.g. pore size, controlling the capillary shrinkage\cite{Brinker1990},
\item the presence of surfactants possibly reducing the capillary pressure\cite{Brinker1990,Kowalski2012},
\item the friction on the substrate\cite{Groisman1994},
\item the drying kinetics,
\item the nature of the solvent flowing though the porous matrix\cite{Brinker1990}\cite{Boulogne2012},
\item the mechanical properties of the porous matrix\cite{Pauchard2009}, by this way, the interaction between particles\cite{Pauchard1999}, the particles toughness, their size, polydispersity affecting the elasto-plastic viscoelasticity.
\end{itemize}

In the following we focus particularly on the effect of \textit{(i)} the drying rate, \textit{(ii)} the nature of the solvent in the  porous media, and \textit{(iii)} the mechanical properties of the colloidal particles on the resulting crack patterns.

\section{Materials and methods}
\label{experimental}

The first system is an aqueous suspension made of silica particles, Ludox HS-40 purchased from Sigma-Aldrich.
The radius of the particles is $9 \pm 2$ nm and the density $2.36\times 10^3$ kg/m$^3$.
The mass fraction of the initial dispersion was estimated to $40.5$\% by a dry extract.
The $pH$ is about $9.0$, the particle surface bears a high negative charge density and DLVO (Derjaguin, Landau, Verwey and Overbeek) theory is expected to apply.

Two different cosolvents are added to the aqueous dispersion: glycerol (purity: $99.0$\%, from Merck) and glucose (purity: $99.5$\%, from Sigma).
These solvents exhibit high miscibility properties in water and low volatility\cite{Daubert1989,Neely1985} compared to water;
thus, we assume glycerol and glucose as non-volatile compounds.
In addition, these cosolvents exhibit highly different intrinsic diffusion coefficients with that of water (intrinsic diffusion coefficient of glycerol in water $\sim 50 \times$ intrinsic diffusion coefficient of water in glycerol\cite{Derrico2004}).
Cosolvent/water weight ratio is prepared using the following process\cite{Boulogne2012}: firstly, a quantity of water is evaporated from the aqueous dispersion by stirring and heating at $35^\circ$C in an Erlenmeyer.
The resulting solution is filtered to $0.5$ $\mu$m to eliminate eventual aggregates; the mass fraction is estimated by a dry extract at $140^\circ$C.
Then, the mass of water removed is replaced by the same mass of the cosolvent.
We note $\kappa$ the ratio between the mass of cosolvent added and the mass of the resulting dispersion.

The second system is aqueous dispersions of nanolatex particles, stable without evaporation, provided by Rhodia Recherche, Aubervilliers, France.
Two different types of particles are considered: rigid ones and deformable ones with two different glass transition temperatures, $T_{g}$.
At room temperature, rigid particles are characterized by $T_{g} = 100^\circ$C and soft particles by $T_{g} = 0^\circ$C; consequently the last can deform more easily than rigid particles.
The  radius of the particles is $15$ nm and the density of pure dry material of rigid and soft particles are respectively $1.08\times 10^3$ kg/m$^3$ and $1.10\times 10^3$ kg/m$^3$.
The volume fraction of both dispersions is $30\%$.
The proportion $\phi_{s}$ in soft particles is defined as $\phi_{s} = V_{s} /(V_{s} + V_{h})$, where $V_{s}$, $V_{h}$ are the volume of soft and rigid particles, respectively.
Binary mixtures are magnetically stirred during 15 min at room temperature, and then sonicated at a $1$ s on/off pulse interval during $300$ s.

\begin{figure}
\centering
\includegraphics[scale=0.5]{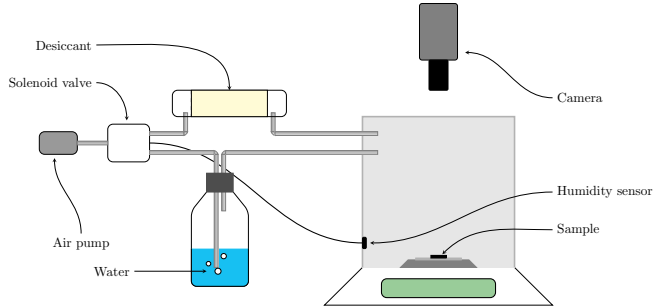}
\caption{Setup.
Dry or moist air is produced by an air flow from the ambient atmosphere through dessicant or water respectively to the scale box.
Depending on the humidity captured by the humidity sensor inside the scale, a solenoid valve is actioned to converge the humidity to the desired value.
}
\label{fig:setup}       
\end{figure}

A circular container (inner diameter $\sim15$ mm, height precised below) is filled with a given amount of dispersion (sample in Figure~\ref{fig:setup}).
The contact line of the dispersion is quenched at the upper edge of the wall and remains pinned all along the drying process.
At the final stage, the layer close to the border of the container exhibits a thickness gradient and results in radial cracks.
On the contrary, in the center of the container we obtain a layer of approximately constant thickness.
In this region, covering about $70\%$ of the total surface area, the evaporation is assumed to be uniform: measurements of crack spacing are investigated there.
The substrate is a non-porous glass plate, carefully cleaned with pure water then ethanol before being dried in a heat chamber at $100^\circ$C.

\begin{figure}
\centering
\includegraphics[scale=0.65]{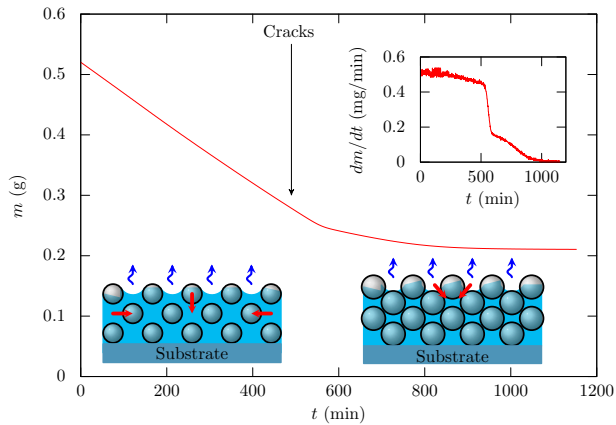}
\caption{Measurements of a sample mass, $m$, during the drying process of a dispersion of Ludox HS-40: the arrow indicates the time for crack formation. Inset: mass variations with time, $\frac{dm}{dt}$, calculated from the mass measurements. Sketch at left: colloidal film during the drying;  sketch at right: solid network saturated with solvent (curvature of the solvent/air menisci occurs at the evaporation surface during solvent loss).}
\label{fig:masse}       
\end{figure}

In our experiments, the transfer of water in air is limited by diffusion and therefore controlled by the relative humidity, $RH$, using a homemade PID controller, at room temperature (Figure \ref{fig:setup}).
The drying kinetics is obtained using a scale (Sartorius) with a precision of $0.01$ mg.
The drying rate is deduced from mass loss as a function of time (Figure \ref{fig:masse}).
At the final stage of the drying process, the  material is still transparent which allows us to observe easily the dynamics of crack patterns in the layer.
The crack morphology is recorded with a camera positioned on the top of the sample.

The macroscopic elastic response of the colloidal  material is characterized using the CSM Instruments Micro Indentation testing (MHT).
The indenter tip (Vickers) is driven into the sample by applying an external force, constant in time.
The penetration depth is recorded as a function of time.

\section{Formation of a crack network in the plane of the layer}

Different stages take place during the drying process of a colloidal film\cite{Coussot2000}.
In a first stage, the evaporation rate is constant, close to the evaporation rate of pure solvent, and hence mainly controlled by the external conditions of relative humidity and temperature in the surroundings.
This stage is usually named Constant Rate Period.
During this regime, the density of particles increases until the formation of a close packing network saturated with solvent (Figure \ref{fig:masse}).
The boundary conditions imposed by both the interface of the film with the substrate and the interface of the film with the air result in a capillary shrinkage at the top layer limited by the adhesion on the substrate that prevents the  material from shrinking\cite{Brinker1990,Xu2009}.
This results in high tension that progressively builds up in the film.

The maximum capillary pressure in the  porous media can reach $P_{cap} = \frac{a \gamma_{s,a} cos(\theta)}{r_{p}}$, where $\gamma_{s,a}$ is the solvent-air surface tension, $\theta$ the liquid/solid contact angle, $r_{p}$ the pore radius and $a$ is a geometrical constant ($\sim 10$)\cite{Brinker1990}.
As a result the tension in the liquid pore compresses the matrix and induces flow from the bulk to the interface.
In the first stage of the drying process, for the meniscus to remain at the surface of the  porous media, the evaporation rate $V_{E}$ has to be equal the flux of liquid $J_{flow}$ at the surface, in accordance with the Darcy law:

\begin{equation}
\label{e.1}
J_{flow} = \frac{k}{\eta}\nabla P\mid_{surface} = V_{E}
\end{equation}

where $k$ is the  porous media permeability, $\eta$ is the viscosity of the liquid and $\nabla P$ the pressure gradient in the liquid pore.

\begin{figure}
\centering
\includegraphics[scale=0.55]{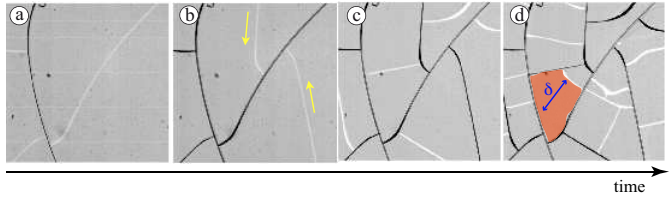}
\caption{Photographs of the formation of a crack pattern. The geometrical orders of the cracks are shown from left to right: first, second, third and fourth order. The cracks of lower orders are drawn in black. In $b$ the arrows highlight the connection of the cracks in white to the existing crack in black. In $c$ the typical size $\delta$ is defined as the square root of the surface area of the fragment colored in red. Photograph size $\sim 500\mu$m.}
\label{fig:hierarchy}       
\end{figure}

When the drying stress reaches a threshold value, cracks form and invade the material.
It takes place at the final stage of the Constant Rate Period, at time $t_{cracking}$ (see Figure \ref{fig:masse}).
The crack network is formed hierarchically: the successive generations of cracks are shown in Figure \ref{fig:hierarchy}.
Finally, cracks delimit adjacent fragments of typical size $\delta$, defined as the square root of the surface area of the fragment (Figure \ref{fig:hierarchy}d).
At the final stage (Figure \ref{fig:hierarchy}d), the pattern is related to the history of the formation since the first crack is the longest, the second crack generation connects to the first one and so on...
Consequently the temporal order of the successive crack generations can be approximately reconstructed \cite{Bohn2005a}.

\section{Effect of the drying rate on the crack patterns}

\begin{figure}
\centering
\includegraphics[scale=0.35]{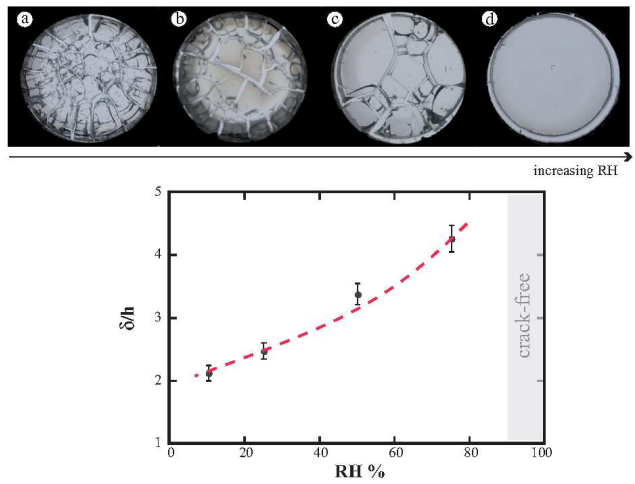}
\caption{Final crack patterns in layers of thickness $h=2$ mm of Ludox HS-40 dried at different humidity rates, $RH$: (a) $RH =25\%$; (b) $RH = 50\%$; (c) $RH = 75\%$; (d) $RH = 95\%$ (the layer is crack-free and a de-adhesion process can be observed between the colloidal material and the border of the circular container).
Graph: ratio between the crack spacing, $\delta$, and the final layer thickness, $h$, plotted as a function of $RH$ ($T = 23^\circ$C; the initial weight deposited in each trough is the same); dashed line is a guide for the eye.}
\label{fig:rh}       
\end{figure}

An amount of colloidal dispersion (Ludox-HS40) is deposited in a circular container and dries at room temperature under a relative humidity $RH$.
This experiment is repeated at different $RH$.
Images in Figure \ref{fig:rh} show the final aspect of the films.
For films dried at low relative humidity numerous cracks are observable in relation to the formation of successive generations of cracks (Figure \ref{fig:rh}a).
For films dried at higher relative humidity, the number of cracks is reduced at the final stage of the drying; consequently, for increasing $RH$ the crack spacing increases as shown in the graph in Figure \ref{fig:rh}.
Moreover drying process at $RH \sim 95\%$ results in a crack-free film (Figure \ref{fig:rh}d).

Under our experimental conditions (the transfer of water in air is limited by diffusion) and during the Constant Rate Period, the evaporation rate depends on the relative humidity as follows:

\begin{equation}
\label{e.2}
V_{E} = D_{w}\frac{1}{L}\frac{n_{wsat}}{n_{1}} (1-(RH/100))
\end{equation}

where $D_{w} = 2.6\times10^{-5} m^2.s^{-1}$ is the diffusion coefficient of water into air, $n_{wsat}$ is the water concentration in air at equilibrium with the film (practically $n_{wsat}$ is almost the same as for pure water, $0.91mol/m^3$), $n_{1}$ the number of water mole per unit volume in liquid water ($55000mol/m^3$).
The length $L$ corresponds to the typical length of the vapor pressure gradient \cite{Dufresne2003}. In our geometry, this length is approximately given by the container radius since the vapor concentration profile is like a hemispherical cap sitting on the container.
The drying timescale is defined as:

\begin{equation}
\label{e.3}
t_{D} = \frac{h}{V_{E}}
\end{equation}

\begin{figure}
\centering
\includegraphics[scale=0.7]{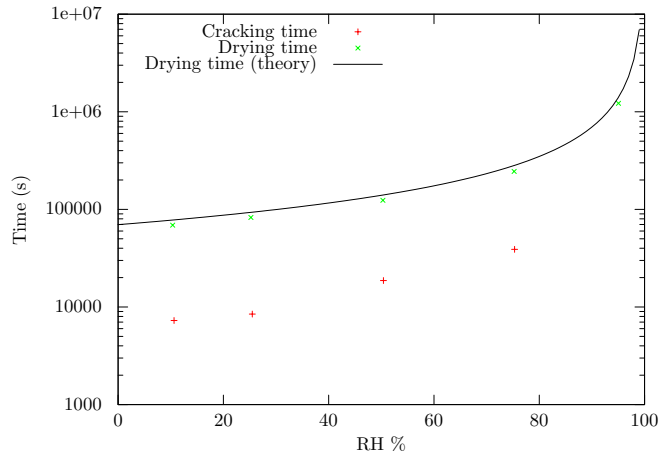}
\caption{Characteristic drying time, calculated from $RH$, and characteristic cracking (measurements) time as a function of the relative humidity, $RH$ for Ludox HS-40 (initial thickness $h=2$ mm). The black line is the theoretical drying timescale from equations \ref{e.2} and \ref{e.3}.}
\label{fig:times}       
\end{figure}

Time $t_{D}$ gives an order of magnitude of the time needed to dry completely a pure solvent film of thickness $h$.
Variations of $t_{D}$ from equation (\ref{e.3}) and $t_{cracking}$, obtained from measurements, are plotted as a function of $RH$ in Figure \ref{fig:times}.
For high $RH$, the evaporation rate is low, the stress developpment is slow resulting in a long period before cracking.

In the following we consider drying conditions at room temperature and $RH\sim50\%$.

\section{Effect of the nature of the solvent on the crack patterns}

\begin{figure}
\centering
\includegraphics[scale=0.35]{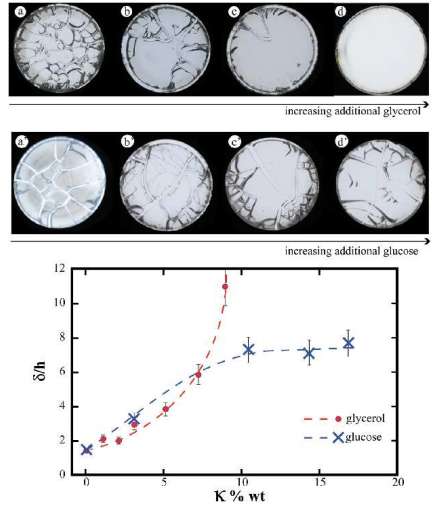}
\caption{Final crack patterns in  films in presence of additional cosolvent to the dispersion of Ludox HS-40 (initial thickness $h=2$ mm).
The cosolvent is glycerol in a,b,c,d : (a) Without any additional glycerol; (b) $\kappa = 5.1\%$; (c) $\kappa = 8.9\%$; (d) $\kappa = 10.2\%$ (the layer is crack-free).
The cosolvent is glucose in a',b',c',d' : (a')  $\kappa = 3.1\%$ ; (b') $\kappa = 10.4\%$; (c') $\kappa = 14.3\%$; (d') $\kappa = 16.8\%$.
Graph: ratio between the crack spacing, $\delta$, and the final layer thickness, $h$, plotted as a function of the concentration in glycerol and in glucose (drying conditions $RH = 50\%$ and $T = 23^\circ$C; the initial weight deposited in each trough is the same); dashed lines are a guide for the eye.}
\label{fig:cosolvents}       
\end{figure}

The addition of a cosolvent to the colloidal dispersion of Ludox HS-40 results, during the drying process, in a gel saturated by a binary mixture of solvent.
The presence of a content of glycerol reduces the number of cracks (Figure \ref{fig:cosolvents} a, b, c, d).
Above a critical concentration in glycerol close to $10$ \%, we obtained a crack-free material (Figure \ref{fig:cosolvents} d).

Similarly, the addition of glucose increases the crack spacing (Figure \ref{fig:cosolvents}).
However, with glucose cracks are not suppressed as glycerol does for concentrations higher than $10\%$.
The crack spacing seems to tend to a maximum value.
Our results indicate that there might be chemical effects (hydrogen bonds) between polyols and silanol groups covering the surface of silica particles.
Gulley \textit{et al.}\cite{Gulley2001} measured the aggregation rate of silica particles with polyols.
They reported that the shorter the polyol molecule is, the higher the stabilizing effect is.
They also remarked that polysaccharides are not stabilizer.
They suggested that, due to the chemical structure, these molecules coat differently the surface of colloids, until the molecules dessorb due to the weakness of the hydrogen bonds and the high pressure between particles.

From the recording of crack dynamics, we remarked that both evaporation time $t_{D}$, and cracking time $t_{cracking}$, are not significantly modified by the presence of the cosolvent (glycerol or glucose).

\section{Effect of the mechanical properties of the particles network on the crack patterns}

\begin{figure}
\centering
\includegraphics[scale=0.35]{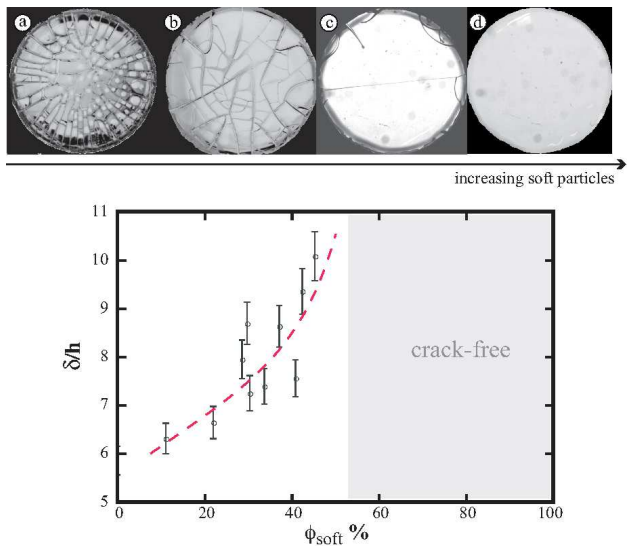}
\caption{Final crack patterns in  films made of both rigid and soft nanoparticles (initial thickness $h=0.5$ mm).
a, b, c, d: the proportion $\phi_{s}$ in soft particles: (a) $\phi_{s} = 0$ ; (a) $\phi_{s} = 40\%$ ; (a) $\phi_{s} = 50\%$ ; (a) $\phi_{s} = 60\%$ (the layer is crack-free).
Graph: ratio between the crack spacing, $\delta$, and the final layer thickness, $h$, plotted as a function of the proportion $\phi_{s}$ in soft particles (drying conditions $RH = 50\%$ and $T = 23^\circ$C; the initial weight deposited in each trough is the same); dashed line is a guide for the eye.}
\label{fig:softhard}       
\end{figure}

The global mechanical properties of the  material can be modified by changing the mechanical properties of the particles themselves.
Final films made of rigid particles are found to contain a large number of cracks as shown previously.
On the contrary, films made of soft particles may be homogeneous\cite{Provder1996}.
Mixtures of rigid and soft particules results in various crack pattern depending on the composition.
Images in Figure \ref{fig:softhard} show final aspect of films composed with various volume fractions in soft particles.
Above a threshold proportion in soft particles, the final film is crack-free (Figure \ref{fig:softhard}d).


\section{Discussion}

The crack formation is due to a mismatch between the mechanical stress in the solid and its strength.
Indeed, if drying stresses exceed the strength of the porous material, they will release in the formation of cracks.
In particular, the drying rate, the effect of the nature of the solvent and the effect of the mechanical properties of the particles network easily result in a modification of the macroscopic response of the solid and so the critical stress for fracture.
The expression of the critical stress derived from the model by Tirumkudulu et al. \cite{Tirumkudulu2005} is:
$\sigma_{c} \sim \bar{G}^{1/3}\big(\frac{\gamma}{h}\big)^{2/3}$, where $\gamma$ is the surface energy solid/air, and $\bar{G}$ corresponds to the macroscopic response of the solid film that expresses as:
$\bar{G} \sim \phi_{m}M G$, here $G$ is the shear modulus of the particles and $\phi_{m}$ is the final close-packed volume fraction and $M$ is the coordination number of the network.

\begin{itemize}

\item Drying rate can affect the organization of the particles network (particularly the coordination number $M$).
We suggest that a slow drying process of the porous material results in a structure capable of supporting the mechanical drying stresses more efficiently.
In this assumption a low drying rate could increase the resistance to cracking.
In addition, a slow drying rate tends to reduce the tension in the liquid near the drying surface of the  layer in accordance with equation (\ref{e.1}).
Consequently the tendency to crack is reduced as shown in section 4.

\item The presence of an additional cosolvent has two consequences on the porous material: a chemical effect and a physical one.
The last effect is due to a combination between the flow driven by the pressure gradient (Darcy law) and diffusion mechanisms (Fick law) in accordance with Scherer work (1989)\cite{Scherer1989} this combination of processes affects the drying process.
Indeed the pressure gradient in the liquid pore, $\nabla P$, is a key parameter in the development of the tensile stress distribution in the porous material.
Indeed, if the pressure in the liquid pore was uniform, the shrinkage of the porous matrix would be uniform too and cracking process were inhibited.
Inversely, the steeper $\nabla P$ is, the greater the difference in shrinkage rate is between the interior and the exterior of the  layer (Figure \ref{fig:sketch-P}).
It results in the flattening of the pressure gradient in the liquid pore (Figure \ref{fig:sketch-P})\cite{Boulogne2012}.
The distribution of the drying stress is more uniform and consequently the cracks are inhibited as shown in section 5.

\begin{figure}
\centering
\includegraphics[scale=0.3]{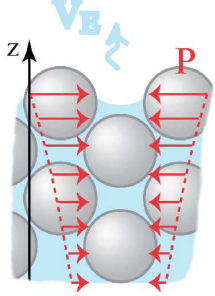}
\caption{Sketch of the pressure gradient in liquid pores.}
\label{fig:sketch-P}       
\end{figure}

\item In the case of a network made of mixtures of rigid and soft particles the macroscopic response of the material is function of the shear modulus of rigid $G_{r}$ and soft $G_{s}$ particles: $\bar{G} = f(G_{r},G_{s})$\cite{Pauchard2009}.
The stress can be released by internal modification of the  structure.
The presence of soft particles releases a part of this stress, consequently reduces the crack formation\cite{Pauchard2009}, as shown in section 6.
\end{itemize}

\begin{figure}
\centering
\includegraphics[scale=0.3]{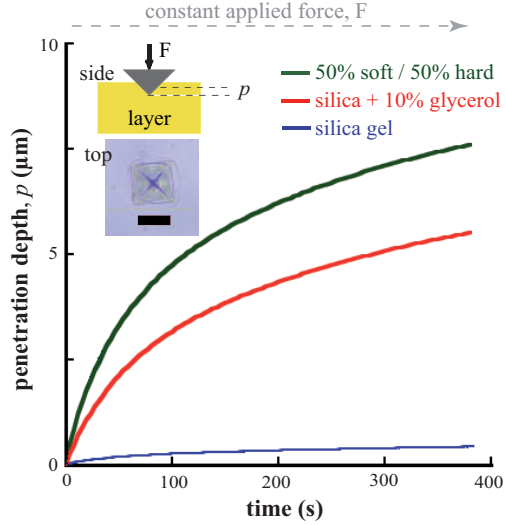}
\caption{Creep comparison of different  dryied layers: the change in depth, $p$, of the indenter tip (Vickers tip) was measured as a function of time from $F=50$ mN indents reached with $50$ mN/min loading rate. Three systems are investigated:  silica Ludox HS-40,  silica Ludox HS-40 with a cosolvent content added, binary mixture of soft nanolatex and rigid nanolatex particles. For each layer the drying process was investigated at $RH=50\%$. Insets: sketch of the indentation testing and print let at the surface of the  layer after removing the indenter tip (bar=$25\mu$m).}
\label{fig:indentation}       
\end{figure}

Moreover the presence of an additional cosolvent or soft particles in the material modifies its mechanical properties.
The mechanical properties are usually characterized by (i) the instantaneous response of an applied force and (ii) the time-dependence of the stress release of the  layer.
Orders of magnitude of both parameters can be investigated using indentation testing.
Case (i) is related to elastic modulus measurement and case (ii) is obtained by creep measurements.
The most common method to measure creep behavior of a material is to maintain the applied force at a constant maximum value, $F$, and measure the change in depth of the indenter, $p$, as a function of time.
Figure \ref{fig:indentation} gives a comparison of creep behaviors for three films.

Modeling the layer by a Kelvin-Voigt two-element model (a purely viscous damper and purely elastic spring connected in parallel), the creep response to an external force $F$ could be expressed as:

\begin{equation}
\label{e.4}
\delta^2(t)=\frac{\pi cot(\alpha)}{2} \frac{F}{E}\left[1-e^{\left(\frac{-t}{\tau}\right)}\right]
\end{equation}

where $\alpha$ is the indenter cone semi-angle, $E$ and $\tau$ are fit parameters representing an elastic modulus of the material (corresponding to the spring element), and a constant that quantifies the time dependent property of the material, respectively.
The values for the two parameters are given in Table \ref{table}.

Accordingly with Figure \ref{fig:indentation}, a layer of an aqueous  suspension made of rigid particles is stiffer than a mixture of soft and hard particles.
Also, in the same way the relaxation time of the material under an external force is larger.

The characteristic time $\tau$ marks the transition from the viscous to the elastic regime.
The larger this relaxation time is, the longer the matrix can be reorganized to relax stresses.
This process competes with the relaxation by the formation of cracks.
As a result, from the data in Table \ref{table}, we can argue that the addition of glycerol or soft particles enhance stresses relaxation leading to less cracks in the dry material.

\begin{table}
\begin{center}
\begin{tabular}{|p{5cm}|c|c|}
\hline
System & $E$ (MPa) & $\tau$ (s)\\
\hline

$50\%$ soft / $50\%$ rigid &  $5 \pm 2$ & $1210$\\\hline
silica  particles in\\ water-glycerol mixture ($90$\%/$10$\%)  &  $40 \pm 8$ & $825$\\\hline
 silica particles in pure water & $100 \pm 110$ & $120$\\\hline
\end{tabular}
\end{center}
\caption{Creep behavior modeled by a Kelvin-Voigt two-element: $\delta^2(t)=\frac{\pi cot(\alpha)}{2} \frac{F}{E}[1-e^{(\frac{-t}{\tau})}]$, where $\alpha = 68^\circ$ and $F = 0.05$ N.}
\label{table}
\end{table}

\section{Conclusion}

During the drying of colloidal suspensions via the pressure gradient in liquid pores, cracks propagate in the  material due to two antagonist mechanisms: the retraction induced by the loss of solvent which is limited by the adhesion of the  layer on the substrate.
These two mechanisms are responsible to the build up of a tensile stress in the material.

In this article, we point out that the resulting crack patterns are affected by various parameters and we studied three of them: drying rate throw the relative humidity, the effect of the solvent volatility and the presence of soft particles among rigid ones.
Using silica nanoparticles in water, a strong dependance of the crack spacing has been observed with the drying kinetic.
Thus, a control of the drying conditions (relative humidity, temperature) is request to ensure reproducible observations.
The volatility of the solvent has been investigated by adding non-volatile compounds (polyols).
As water evaporates, mixing Fick's fluxes emerge in the porous medium resulting in a flattened pressure gradient and an increase of the crack spacing.
Moreover, the crack spacing is quantitatively affected by the non-volatile compound.
Molecular structures are believed to modify silanol-polyol interactions.
Finally, we also observed a reduction of cracks in mixtures of soft and hard particles: soft particles are allowed to deform during the drying, storing the mechanical energy.

For both later systems, we check that kinetic is not affected by the addition of an extra compound, but the caracteristic relaxation time is significantly increased.
This explains a reduction of the tensile stress leading to a decrease of the number of cracks.

Futher investigations on the drying in presence of molecules in the solvent are necessary for a better understanding of the state of final materials.
Especially, it would be interesting to study the influence of molecular structures in order to tune the adsoprtion-desorption energy.

\section{Acknowledgment}
The authors thank Mikolaj Swider for his participation to the Relative Humidity experiment and Alban Aubertin for designing the PID controller.
%
\bibliography{biblio}
\bibliographystyle{plain}

\end{document}